\documentclass[%
 preprint, 
 superscriptaddress, longbibliography, amsmath, amssymb, aps, prl, floatfix,
 ]{revtex4-2}

\usepackage{graphicx}
\usepackage{dcolumn}
\usepackage{bm}
\usepackage[colorlinks=true,linkcolor=blue,citecolor=blue,urlcolor=blue]{hyperref}
\newcommand{\MoS}{MoS$_2$}
\newcommand{\Kp}{\ensuremath{K'}}
\newcommand{\sigp}{\ensuremath{\sigma^{+}}}
\newcommand{\sigm}{\ensuremath{\sigma^{-}}}
\newcommand{\dmu}{\ensuremath{\Delta\mu}}
\newcommand{\CD}{\ensuremath{\mathrm{CD}}}
\newcommand{\VP}{\ensuremath{\mathrm{VP}}}
\newcommand{\dpm}{\ensuremath{d_{\pm2}}}

\newcommand{\Ttwo}{\ensuremath{T_2}}
\newcommand{\GWcm}{\ensuremath{\mathrm{GW/cm^2}}}
\newcommand{\ev}[1]{\ensuremath{#1~\mathrm{eV}}}
\newcommand{\fs}[1]{\ensuremath{#1~\mathrm{fs}}}

\graphicspath{{figures/}}

\begin{document}


\title{Carrier-Resolved Attosecond Valley Polarimetry of Monolayer MoS$_2$}

\author{Navdeep Rana}
\email[Contact author: ]{navdeeprana@lsu.edu}
\affiliation{Department of Physics and Astronomy, Louisiana State University, Baton Rouge, Louisiana 70803, USA}

\author{Lun Yue}
\email{lyue2@binghamton.edu}
\affiliation{Department of Physics, Binghamton University, Binghamton, New York 13902, USA}

\author{Mette B. Gaarde}
\email{mgaarde1@lsu.edu}
\affiliation{Department of Physics and Astronomy, Louisiana State University, Baton Rouge, Louisiana 70803, USA}

\date{\today}

\begin{abstract}
In recent years, all-optical writing and switching of valley polarization in two-dimensional semiconductors has been demonstrated on femtosecond timescales. Reading this polarization out on the same timescale, carrier by carrier, has so far remained out of reach. Employing semiconductor Bloch equation simulations of monolayer MoS$_2$, we show that attosecond transient absorption closes this gap by turning the Mo $4p$ semicore edge into a quantitative valley polarimeter. We find  that pump-enabled core-to-valence absorption probes the holes while core-to-conduction bleaching probes the electrons, so that a single spectrum identifies each carrier by its photoabsorption energy. The hole channel, Pauli-blocked in equilibrium, emerges background-free. Its circular dichroism reverses sign with the pump helicity, and its normalized magnitude is proportional to the valley polarization. Finally, we show that scanning the probe delay clocks the few-femtosecond write in real time. Thus, attosecond core-level dichroism is a carrier-sensitive, quantitative probe of the creation and evolution of valley polarization.
\end{abstract}

\maketitle

The valley pseudospin of two-dimensional (2D) hexagonal semiconductors is a time-reversal-protected information carrier, locked to spin in monolayer transition-metal dichalcogenides (TMDs)~\cite{xiao2007valley,xiao2012coupled,xu2014spin}. Encoding information in the valley index is the goal of valleytronics, which envisions devices that write, switch, and read out this information on the ultrafast timescales of light~\cite{schaibley2016valleytronics,vitale2018valleytronics}. The ability to {\it write} valley polarization (\VP) is now well established through resonant circular excitation~\cite{mak2012control,zeng2012valley,cao2012valley}, sub-cycle tailored fields~\cite{jimenez2020lightwave,jimenez2021sub,mrudul2021light}, and transient symmetry engineering~\cite{tyulnev2024valleytronics,mitra2024light}. All-optical {\it switching} of the written polarization has been shown as well, from sub-cycle lightwave control~\cite{langer2018lightwave} to coherent pulse protocols~\cite{rana2023all,silva2022all,gucci2026encoding}. By contrast, methods to {\it read} the valley state have progressed more slowly. The recent valleytronics roadmap identifies fast, quantitative readout as a bottleneck for the field~\cite{seyler2026valleytronics}. 

Existing readouts span picoseconds to attoseconds, but each trades away carrier resolution, direct access to populations, or a quantitative calibration. Polarization-resolved photoluminescence and Kerr rotation read \VP{} indirectly, through its excitonic signatures, but too slowly to follow how it is written~\cite{mak2012control,zeng2012valley,dal2015ultrafast}. Harmonic-generation and high-order sideband techniques extend the time resolution into the femtosecond regime but infer the valley state from the emitted spectrum rather than measuring the underlying populations~\cite{langer2018lightwave,mrudul2021light}. Most recently, Li {\it et al.}~\cite{li2025attosecond} showed that the circular dichroism (CD) of inter-conduction-band transitions can track \VP{} with 250-as resolution. The observable in Ref.~\cite{li2025attosecond}, however, is a single scalar, the net electron imbalance in the lowest conduction band. The holes, which carry the same valley index, are not accessed. Converting the imbalance into a valley polarization requires the total excited population, which changes with every pump condition and must be supplied by an independent measurement such as time- and angle-resolved photoemission.
\begin{figure}[t]
  \centering
  \includegraphics[width=\columnwidth]{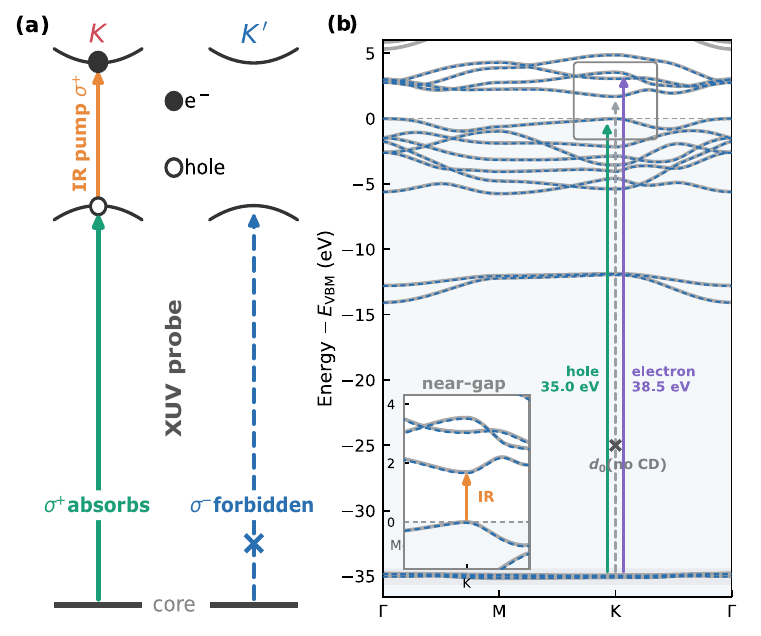}
  \caption{Core-level valley readout in monolayer \MoS{}. (a)~Readout scheme for a \sigp{} pump at $K$. The pump (orange) creates an electron--hole pair in the $K$ valley. The co-rotating \sigp{} XUV probe (green) is  absorbed on the core$\to$valence transition. The counter-rotating \sigm{} probe (blue, dashed) addresses \Kp{}, where the filled valence band blocks this transition. The hole readout is therefore background-free. (b)~Wannier bands (dashed) compared with PBE-DFT bands (gray). The flat Mo~$4p$ semicore couples to the band extrema through two spectrally separated channels. The hole channel lies at \ev{35.0} and the electron channel near \ev{38.5}. Transitions into the $d_0$ conduction minimum (cross) carry no CD. The inset magnifies the boxed region around the direct gap at $K$. The orange arrow marks the resonant IR pump (\ev{1.69}). }
  \label{fig:concept}
\end{figure}

In this Letter, we show that probing a {\it core} level via attosecond transient absorption (ATA) turns the energy-resolved CD into a carrier-resolved valley polarimeter [Fig.~\ref{fig:concept}(a)]. In monolayer MoS$_2$, the Mo $4p$ semicore level lies $\approx\ev{35}$ below the valence-band maximum and is nearly dispersionless across the Brillouin zone. A single attosecond pulse spanning this edge simultaneously accesses transitions to both the valence and conduction bands, and we show that the transition energy alone identifies the carrier [Fig.~\ref{fig:concept}(b)]. The core$\to$valence transition at \ev{35.0} is Pauli blocked in equilibrium and appears only where optical excitation has created holes. The hole population can therefore be measured against a dark background. At the same time, core$\to$conduction transitions near \ev{38.5} probe the excited electrons. The valley selectivity follows from the $C_3$ crystal-angular-momentum selection rules. The valence-band (VB) maximum and the conduction bands (CBs) probed near \ev{38.5} absorb light of opposite helicity in the two valleys, whereas the lowest CB, of $d_{z^2}$ character, responds to both helicities equally and is therefore valley-blind. Our calculations show that these selection rules provide a quantitative readout of the written valley polarization. The hole-line CD reverses with the pump helicity and vanishes for a  linearly polarized pump, as required by time-reversal symmetry. We show that across both ellipticity and pump-intensity scans, the dichroic signal remains proportional to the valley population imbalance, so the measured spectrum converts directly into an absolute valley polarization. Finally, by scanning the probe delay, we show that the signal follows the buildup of the hole population, resolving the 4-fs writing process in real time.

Our results provide a time-resolved demonstration of the core-level valley selectivity predicted from crystal symmetry~\cite{geondzhian2022valley}. That work showed that circularly polarized core$\to$valence absorption in TMDs becomes valley-selective once the valence-band maximum is depleted, for instance by optical excitation, whereas the core$\to$conduction-band minimum remains valley-blind. Here, the optical pump creates the required hole population, and the attosecond probe extends this selectivity into the time domain. We also note that, because the semicore level is localized on the Mo atoms, the probe is element specific and selectively addresses the TMD layer, even in encapsulated or stacked heterostructures where valleytronic devices are realized~\cite{attar2020simultaneous,buades2021attosecond}.

We describe monolayer \MoS{} with a 16-Wannier-function tight-binding model derived from a PBE-DFT calculation. The model spans the Mo $4p$ semicore to the conduction bands $\approx\ev{5}$ above the VB maximum and reproduces the DFT bands throughout [Fig.~\ref{fig:concept}(b)], including the \ev{1.68} direct bandgap at $K$/\Kp{}. The semicore comprises the three spinless Mo $4p$ orbitals, split by \ev{0.16}. Spin--orbit coupling is omitted in the results shown below. In the supplementary material (SM~\cite{SM}) we show that the main conclusions remain unchanged when spin--orbit coupling is included.

We propagate the 16-band model with the semiconductor Bloch equations in the Wannier gauge~\cite{silva2019high}. We apply a global dephasing time $\Ttwo=\fs{10}$ to the interband coherences and neglect population relaxation. The pump is a circularly or elliptically polarized optical pulse resonant with the $K$ gap ($\hbar\omega=\ev{1.69}$ photon energy, \fs{7.2} pulse duration full-width at half-maximum (FWHM), 100~\GWcm{} intensity). The probe is a weak circularly polarized, 400 attosecond pulse centered at \ev{37.5} covering the 33--42~eV window. The pump--probe delay $\tau$ is positive when the probe arrives after the pump center. From the probe-induced current we compute the XUV absorption spectrum of the monolayer with and without the pump. The difference of the two spectra is the transient absorption $\dmu_{\pm}(\omega,\tau)$ for a right- ($\sigma^{+}$) or left-circularly polarized ($\sigma^{-}$) probe at delay $\tau$, the quantity measured in XUV transient-absorption experiments~\cite{attar2020simultaneous,buades2021attosecond}. The CD is the difference between the transient absorption spectra for the two probe helicities, $\CD=\dmu_{+}-\dmu_{-}$. Explicit expressions for $\dmu_{\pm}$ and the complete set of simulation parameters are given in \hyperref[app:methods]{Appendix~A}.

The CD owes its valley sensitivity to the crystal symmetry at $K$ and \Kp{}. There, the little group, the set of operations that map the valley onto itself, is $C_{3h}$~\cite{NoteLittleGroup}. The threefold rotation $\hat{C}_{3}$ assigns each state a crystal angular momentum $m$, and $\sigma^{\pm}$ light connects states with $m_f=m_i\pm1$ (mod 3)~\cite{geondzhian2022valley,cao2012valley,rana2022generation,rana2022probing}. The horizontal mirror $\hat{\sigma}_{h}$ assigns each state a parity that $\sigma^{\pm}$ light cannot change.  The helicity selectivity of a core$\to$band channel is quantified by the asymmetry
\begin{equation}
  \eta=\frac{|D_{+}|^{2}-|D_{-}|^{2}}{|D_{+}|^{2}+|D_{-}|^{2}},
  \label{eq:eta}
\end{equation}
with $D_{\pm}$ the dipole matrix elements for $\sigma^{\pm}$ light. Table~\ref{tab:rules} lists the resulting values for the five channels at $K$ (derivation in \hyperref[app:selection]{Appendix~B}). The \dpm{} states at the VB maximum and in the upper CBs are maximally dichroic ($|\eta|\to1$). Each couples to a single helicity. The $d_0$ CB minimum is the opposite extreme, a symmetry-protected null. Both helicities reach it with equal amplitude, $\sigma^{+}$ from $p_{-}$ and $\sigma^{-}$ from $p_{+}$, so its dichroism vanishes. Time reversal enforces $\eta(\Kp{})=-\eta(K)$, so the {\it sign} of the CD is the valley index while the {\it transition energy} selects the carrier. Throughout we follow the standard convention in which $K$ is the valley addressed by \sigp{} light~\cite{xiao2012coupled}.

\begin{table}[b]
  \caption{Calculated Mo $4p$ core$\to$band transition energies and helicity asymmetries at $K$.  The listed $d_m$ orbitals denote the dominant Mo character of the final states. The parameter $\eta$ is defined in Eq.~(\ref{eq:eta}), and time-reversal symmetry enforces $\eta(K')=-\eta(K)$. The VB-max transition is Pauli blocked in equilibrium and is activated by hole creation, defining the hole channel. Among the conduction channels, CB min provides a near-null helicity reference, whereas CB$+2$ underlies the dominant electron-channel bleach.}
  \label{tab:rules}
  \begin{ruledtabular}
  \begin{tabular}{lccc}
   Final band & $E_{\mathrm{tr}}$ (eV) & Mo $d_m$  & $\eta(K)$ \\
   \colrule
   VB max   & 35.0  & $d_{+2}$   & $+1.00$ \\
   CB min   & 36.7  & $d_{0}$    & $+0.02$ \\
   CB$+1$   & 38.1  & $d_{-1}$   & $-1.00$ \\
   CB$+2$   & 38.6  & $d_{-2}$   & $-1.00$ \\
   CB$+3$   & 39.9  & $d_{+1}$   & $+1.00$ \\
  \end{tabular}
  \end{ruledtabular}
\end{table}

\begin{figure}[t]
  \centering
  \includegraphics[width=\columnwidth]{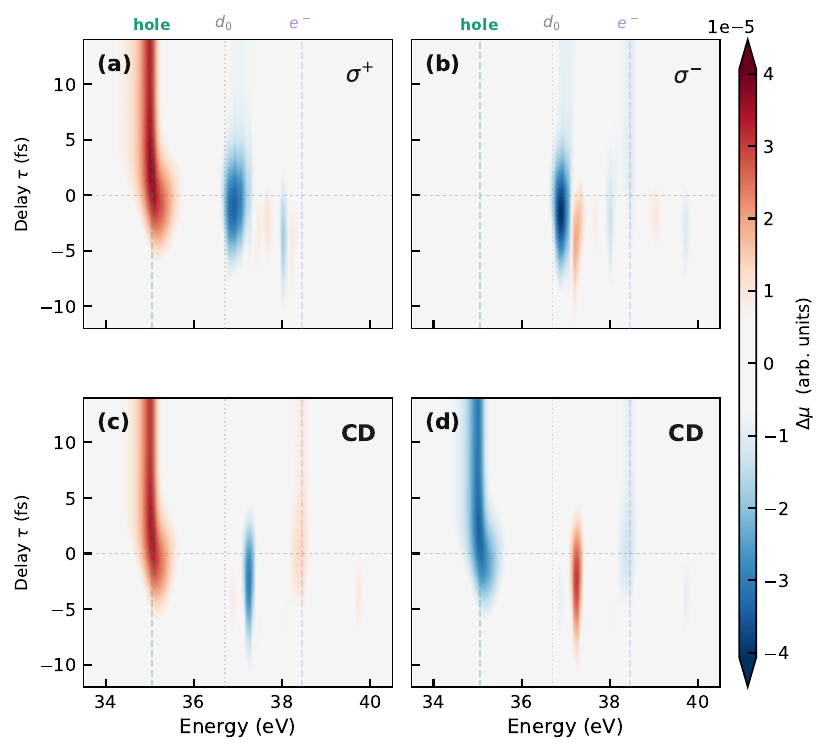}
  \caption{Two-channel, carrier-resolved readout. (a),(b)~Transient absorption $\dmu_{\pm}(\omega,\tau)$ for \sigp{} and \sigm{} probes after a \sigp{} pump, which writes carriers in the $K$ valley. The co-rotating probe (a) reveals the core$\to$valence hole line at \ev{35.0}. For the counter-rotating probe (b), the hole line is dark and the electron channel near \ev{38.5} (dashed line, e$^-$) appears as a bleach of the bright edge. Both probes exhibit a dominant common bleach at 36.7--37.5~eV (dashed line, $d_0$). (c),(d)~Circular dichroism $\CD=\dmu_{+}-\dmu_{-}$ for \sigp{} and \sigm{} pumps, respectively. The CD reverses sign exactly upon reversal of the pump helicity, as required by time-reversal symmetry, and the sign of the hole-line CD directly labels the written valley.}
  \label{fig:readout}
\end{figure}

Figure~\ref{fig:readout} establishes core-level ATA spectroscopy as an all-optical readout of valley polarization. A resonant \sigp{} pump creates electron--hole pairs predominantly in the $K$ valley. Because optical excitation conserves the valley index, each pair resides in its own valley, and a single quantity characterizes the written polarization. We define it as the normalized population imbalance $\VP=(N_{K}-N_{K'})/(N_{K}+N_{K'})$, where $N_{K}$ and $N_{K'}$ are the momentum-resolved carrier populations integrated around each valley~\cite{SM}. The pump writes $\VP=0.92$. Figures~\ref{fig:readout}(a) and \ref{fig:readout}(b) show the transient absorption spectra $\dmu_{\pm}(\omega,\tau)$ for \sigp{} and \sigm{} probes following a \sigp{} pump. The co-rotating \sigp{} probe [Fig.~\ref{fig:readout}(a)] reveals a distinct pump-induced core$\to$valence hole line at \ev{35.0} that is absent before the pump and persists after the pump pulse. In contrast, the counter-rotating \sigm{} probe [Fig.~\ref{fig:readout}(b)] leaves the hole line nearly dark on the same color scale. At a pump--probe delay of \fs{12.9}, well after the pump, the hole-line signal is 17 times stronger for the co-rotating probe than for the counter-rotating one. A weak residual hole-line signal for the \sigm{} probe (not visible with the color scale used in Fig.~\ref{fig:readout}) is a measure of the minority-valley holes, consistent with the written valley polarization of 0.92. We note that during pump-probe overlap, a Stark shift of the hole line appears along with other transient structures.  

An electron channel appears in the \sigm{} spectrum in Fig.~\ref{fig:readout}(b), as a faint bleach (reduced absorption) at \ev{38.45}. This feature originates from pump-induced population in CB$+1$ and CB$+2$  reached via two-photon excitation for a range of momenta around $K$. The two contributions merge into a single bleach rather than appearing as separate CB$+1$ and CB$+2$ lines. The opposite signs of the hole and electron channels follow directly from Pauli blocking. In equilibrium, the filled VB blocks the core$\to$valence transition, while the empty CBs absorb the probe strongly at the bright edge. The pump reverses both. It creates holes and switches the core$\to$valence transition on, so the hole line appears as a positive $\dmu$. It fills CB states and removes part of the edge absorption, so two features appear as a negative $\dmu$. The bleach of the bright edge sets in at the $d_0$ line, and the electron channel sits at the e$^-$ line. We note here that not all transient features are valley selective. The transient structures during temporal pump--probe overlap, the free-induction-decay fringes at negative delay,  and part of the edge bleach are helicity-even and cancel in the CD, which isolates the valley-selective response.

Our calculations show that the CD directly maps the valley polarization onto the attosecond spectrum. Figure~\ref{fig:readout}(c) shows the CD for the \sigp{} pump and Fig.~\ref{fig:readout}(d) for the \sigm{} pump. Reversing the pump helicity inverts the CD at every energy and delay, as required by time-reversal symmetry for a valley-dependent observable. The strongest CD appears at the hole line, while in the electron channels it separates the valley-selective bleach from the overlapping helicity-even edge response. At the conduction-minimum line at \ev{36.7}, the CD is strongly suppressed, consistent with the symmetry-protected null of Table~\ref{tab:rules}. These results establish a carrier- and valley-resolved readout scheme. The photon energy identifies the carrier, with the core$\to$valence absorption at \ev{35.0} reading the holes and the core$\to$conduction bleach near \ev{38.5} reading the electrons. The sign of the CD assigns the valley, positive for a $K$ population and negative for \Kp{}. A single attosecond spectrum therefore separates electrons and holes spectrally and resolves their valley polarization through the CD.

\begin{figure}[t]
  \centering
  \includegraphics[width=\columnwidth]{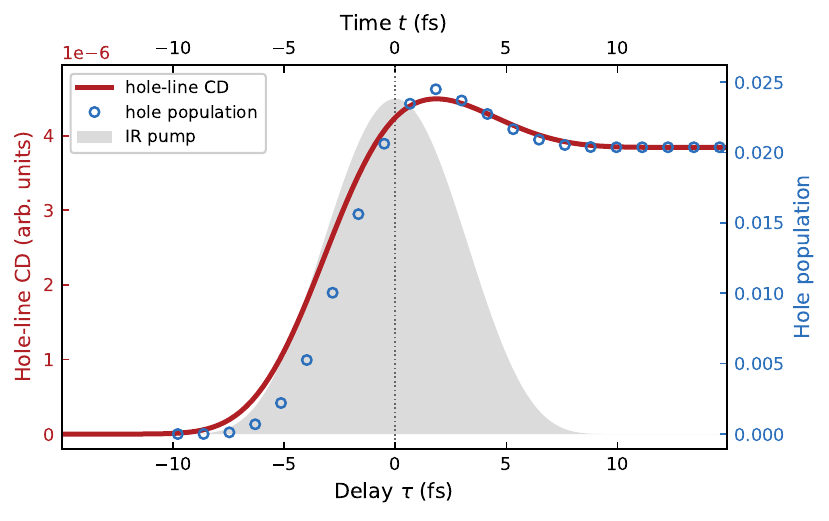}
  \caption{Attosecond readout of valley-polarization buildup (\sigp{} pump). Before the pump, the hole line CD is absent. As the pump arrives (shaded envelope, real time on top axis), the hole-line readout (solid line, left axis) tracks the simulated $K$-valley hole population (circles, right axis). The readout rises from 10\% to 90\% in \fs{4.0}, the population in \fs{4.2}. The short probe samples the population at $t=\tau$.}
  \label{fig:buildup}
\end{figure}

Next, we show that scanning the pump--probe delay resolves the VP as it is written. Figure~\ref{fig:buildup} shows the hole-line CD as a function of pump--probe delay (left axis), together with the simulated $K$-valley hole population (right axis). Before the pump arrives, the hole-line signal is absent. It rises with the pump and tracks the hole population throughout the excitation, with 10--90\% rise times of \fs{4.0} for the readout and \fs{4.2} for the population. Because the \fs{0.4} probe is ten times shorter than this rise, the measured buildup is pump limited. The electron channel exhibits the same buildup~\cite{SM}. The delay axis therefore provides a direct time-resolved readout of the valley write, beyond the reach of exciton-mediated probes and so far proposed only through the conduction-electron imbalance~\cite{li2025attosecond}. 

Another aspect of a practical valley readout is the support of a complete write--read--switch cycle. We have shown the first two operations above, and the polarimeter extends directly to the third. We simulate a three-pulse $\sigma^{+}\sigma^{-}\sigma^{-}$ sequence that writes, cancels, and reverses the net valley polarization. The delayed attosecond probe tracks the reversal of the hole-line CD throughout the sequence, a direct real-time readout of valley switching~\cite{SM}.

\begin{figure}[t]
  \centering
  \includegraphics[width=\columnwidth]{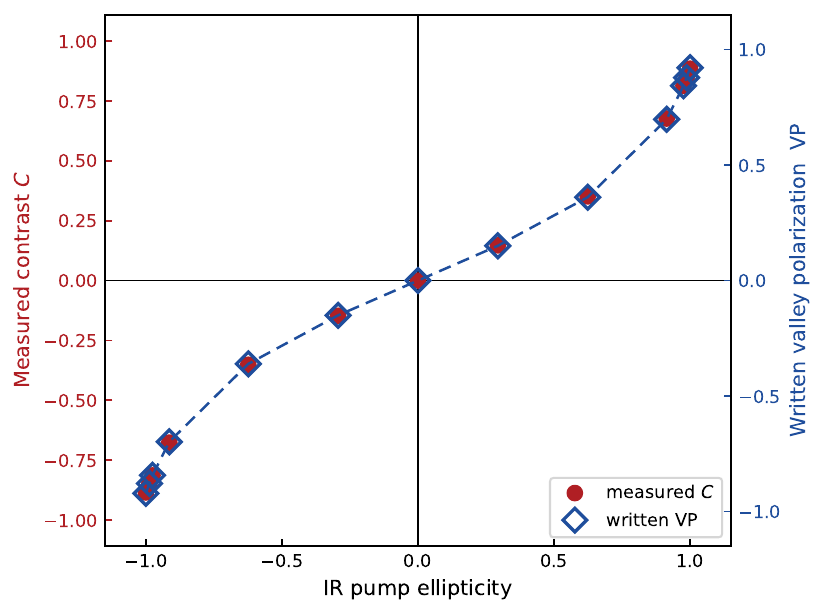}
  \caption{The valley polarimeter. Scanning the pump polarization from \sigm{} through linear polarization to \sigp{} tunes the written \VP{} from $-0.92$ to $+0.92$ (blue diamonds, extracted from momentum-resolved populations). The measured hole-line contrast $C$ (red circles), taken at a pump--probe delay of \fs{12.9}, falls on the same curve. A linear fit yields $C=0.96\cdot\VP$ (Pearson $r>0.9999$). For a linearly polarized pump, both vanish.}
  \label{fig:meter}
\end{figure}

Finally, we establish the hole-line contrast as an absolute valley polarimeter. Figure~\ref{fig:meter} shows the hole-line contrast $C=(\dmu_{+}-\dmu_{-})/(\dmu_{+} + \dmu_{-})$ (left axis, circles) and the written valley polarization (right axis, diamonds) as functions of the pump ellipticity, varied from \sigm{} through linear to \sigp{} at fixed peak intensity. Each ellipticity writes a distinct valley polarization, which we determine from the momentum-resolved carrier populations. The contrast follows the written polarization linearly over the full range, $C=0.96\cdot\VP$, with a Pearson correlation coefficient $r>0.9999$. The slope therefore defines the calibration constant of the polarimeter, while its 4\% deviation from unity reflects the residual cross-valley leakage of the hole channel. Reducing the pump intensity by an order of magnitude changes this calibration by only 1\%, and the complete readout at 10~\GWcm{} is presented in the SM~\cite{SM}. Consequently, a single calibration constant converts the contrast directly into the valley polarization, independent of the pump ellipticity or intensity used to write it. Because the contrast is a ratio, however, it is insensitive to the total carrier population. We recover the population imbalance from the raw CD, which preserves the overall signal amplitude and obeys a second linear calibration, $\CD=\kappa\,(N_{K}-N_{K'})$, with a single proportionality constant $\kappa$ ($r=0.998$) across all ellipticities and both pump intensities. Once the two constants are fixed, by theory or by a one-time reference measurement, a single attosecond spectrum determines both the valley polarization and the absolute carrier population imbalance behind it. The calibration is also predictive. With the leakage fixed by the fitted slope, it gives hole-line on/off contrasts of 17 at 100~\GWcm{} and 49 at 10~\GWcm{}, against the simulated 17 and 50~\cite{SM}.

Importantly, a large transient absorption signal does not necessarily imply valley polarization. A linearly polarized pump still injects electron--hole pairs, and a single-helicity spectrum retains 93\% of the maximum hole-line amplitude. A measurement based on the absorption strength alone could therefore mistake this large carrier population for a valley-polarized state. The CD removes this ambiguity. It collapses by four orders of magnitude at linear polarization and grows continuously with the pump ellipticity (Fig.~\ref{fig:meter}). It thus isolates the time-reversal-odd valley response and rejects the time-reversal-even carrier background. The $d_0$ line [Figs.~\ref{fig:readout}(c) and \ref{fig:readout}(d)] reinforces these checks, since its strongly suppressed dichroism provides an internal reference against instrumental asymmetries. Together, the helicity reversal of the CD, the symmetry protected $d_0$ null, and the suppression of the CD for linearly polarized pumping constitute three independent symmetry tests, and the readout passes all three~\cite{SM}.

In the SM \cite{SM}, we show in detail that the valley polarimeter is robust against spin--orbit coupling. Briefly, in a fully relativistic 32-band model, the Mo $4p$ core splits into a $p_{3/2}$/$p_{1/2}$ doublet separated by \ev{2.46}. Because the selection rules of Table~\ref{tab:rules} act on the orbital part of the wavefunction, the computed hole-channel helicity asymmetry remains near unity for both sub-edges. The hole line shifts to the $p_{3/2}$ sub-edge near \ev{34.3}, where it remains dark for the counter-rotating probe and preserves its 1.6~eV separation from the bright edge. The phase-resolved readout is likewise preserved under spin--orbit coupling~\cite{SM}. Summing over the core multiplet reduces the electron-channel contrast, whereas the hole line remains a robust valley marker~\cite{SM}.

In summary, we have established attosecond circular dichroism at the Mo $4p$ edge of monolayer \MoS{} as a quantitative, carrier-resolved probe of valley polarization. The readout probes electrons and holes separately, and resolves the writing of valley polarization in real time. This capability rests on three features. (i)~It is carrier resolved. Holes and electrons are distinguished by their core-level transition energies within a single element-specific absorption edge. (ii)~It is background-free. The hole line emerges from a Pauli-blocked dark background and is visible directly in a single co-rotating spectrum with an on/off contrast of 17:1, rather than as a differential signal on a bright absorption edge. (iii)~It is calibrated. The linear relation $C=0.96\cdot\VP$ and a single proportionality constant $\kappa$ across excitation conditions define a polarimeter whose calibration is fixed once and then holds across different pumps, with the $d_0$ line and  linear-polarization limit as internal symmetry references. It therefore provides a general route to quantitative measurements of non-equilibrium valley dynamics, with sufficient temporal resolution to follow the creation, switching, and decay of valley polarization separately for electrons and holes.

Finally, the proposed readout is within current experimental reach~\cite{SM}. The pump-induced hole line reaches 3.6\% of the static edge absorption, corresponding to an optical-density change $\Delta\mathrm{OD}\sim10^{-4}$ at the percent-level absorbance of a monolayer, within the sensitivity demonstrated by femtosecond XUV transient absorption at this edge~\cite{attar2020simultaneous,buades2021attosecond} and by attosecond core-exciton spectroscopy in wide-gap films~\cite{gannan2025polarization}. Circularly polarized XUV probes are available from bicircular high-harmonic sources~\cite{fleischer2014spin,kfir2015generation,brooks2024circularly} and polarization-controlled free-electron lasers~\cite{allaria2014control,lutman2016polarization}. A real sample will depart from the model in two ways. Core-hole attraction will reshape the lines into core-exciton resonances~\cite{qiu2013optical}, and carrier scattering will redistribute the written populations within tens of femtoseconds~\cite{attar2020simultaneous}. Neither alters the symmetry basis of the readout, though both reshape the lines and shift the numerical calibration. The selection rules, the nulls, and the proportionality to the written imbalance rest on symmetry and Pauli blocking, not on line shapes, and the redistribution is itself the first target. Watching that decay, carrier by carrier, is what this probe is aimed for.

\begin{acknowledgments}
Work at LSU was supported by the U.S. Department of Energy, Office of Science, Basic Energy Sciences under Contract No. DE-SC001043, and high-performance computational resources were provided by LSU and the Louisiana Optical Network Infrastructure (LONI).
\end{acknowledgments}

\appendix

\section{Appendix A: Methods}
\label{app:methods}

Ground-state electronic structure is computed within PBE density-functional theory (\textsc{Quantum 
ESPRESSO})~\cite{ giannozzi2009quantum,giannozzi2017advanced} using projector-augmented-wave pseudopotentials~\cite{blochl1994projector} that include the Mo $4s$ and $4p$ semicore states in the valence, 120/1200-Ry wave-function/charge-density cutoffs, and a $25\times25\times1$ Monkhorst--Pack $k$ grid. Sixteen maximally localized Wannier functions are disentangled from fifty Bloch bands with \textsc{Wannier90}~\cite{marzari1997maximally,souza2001maximally,pizzi2020wannier90,yates2007spectral,marzari2012maximally} on a $60\times60\times1$ $k$ grid and provide the Hamiltonian for the time propagation.

The time evolution of the density matrix $\rho_{\mathbf{k}}$ is governed by the semiconductor Bloch equations in the Wannier gauge, in atomic units ($\hbar=e=m_e=4\pi\varepsilon_0=1$)~\cite{silva2019high},
\begin{equation}
\begin{split}
  i\,\partial_t \rho_{\mathbf{k}}
  &= \Big[H^{W}_{\mathbf{k}}+\mathbf{E}(t)\cdot\boldsymbol{\mathcal{A}}^{W}_{\mathbf{k}},\,\rho_{\mathbf{k}}\Big] \\
  &\quad + i\,\mathbf{E}(t)\cdot\nabla_{\mathbf{k}}\,\rho_{\mathbf{k}}
  - i\,\mathcal{D}\big[\rho_{\mathbf{k}}\big],
\end{split}
\label{eq:sbe}
\end{equation}
where $H^{W}_{\mathbf{k}}$ and $\boldsymbol{\mathcal{A}}^{W}_{\mathbf{k}}$ denote the Wannier-gauge Hamiltonian and Berry connection, respectively. $\mathbf{E}(t)$ is the combined pump and probe field. The initial state has the twelve occupied bands filled and the four conduction bands empty, and the time propagation uses a fourth-order Runge--Kutta scheme with a $0.1$~a.u.\ time step. The dephasing operator $\mathcal{D}$ acts in the Hamiltonian eigenbasis, damping all interband coherences at the rate $1/\Ttwo$ while leaving the band populations unchanged. We model the dephasing with $\Ttwo=\fs{10}$ by multiplying the interband coherences by $\exp(-\Delta t/\Ttwo)$ at each dephasing step, $\Delta t=0.2$~a.u. On the same time grid, we evaluate the current in the Hamiltonian gauge~\cite{wang2006ab},
\begin{equation}
\begin{split}
  \mathbf{j}(t) &= -\frac{1}{N_k}\sum_{\mathbf{k}}
    \mathrm{Tr}\!\left[\rho^{H}_{\mathbf{k}}(t)\,\mathbf{v}^{H}_{\mathbf{k}}\right], \\
  \mathbf{v}^{H}_{\mathbf{k}} &= \nabla_{\mathbf{k}} H^{H}_{\mathbf{k}}
  + i\big[H^{H}_{\mathbf{k}},\boldsymbol{\mathcal{A}}^{H}_{\mathbf{k}}\big],
\end{split}
\label{eq:current}
\end{equation}
with $N_{\mathbf{k}}$ the number of $k$ points, $\rho^{H}_{\mathbf{k}}=U^{\dagger}_{\mathbf{k}}\rho_{\mathbf{k}}U_{\mathbf{k}}$ the density matrix in the Hamiltonian eigenbasis, and $U_{\mathbf{k}}$ the unitary matrix that diagonalizes $H^{W}_{\mathbf{k}}$. The first term is the intraband contribution and the second the interband contribution, which the calculation evaluates separately~\cite{yue2022introduction}. The total current is used for plotting every spectrum in this work. Table~\ref{tab:pulses} summarizes the pump and probe parameters, with durations quoted as intensity FWHM. Elliptically polarized pumps are constructed by scaling the minor-axis field with a $90^\circ$ phase offset while keeping the major-axis amplitude fixed, which keeps the peak intensity constant while the fluence varies with ellipticity. The ellipticity axis of Fig.~\ref{fig:meter} is the Stokes parameter $S_3$ of the propagated field. Probe delays span $-15$ to $+15$~fs in 43 steps, and unless otherwise specified fixed-delay calculations are performed at $\tau=\fs{12.9}$, well after the pump. Time-reversed partners are propagated independently rather than being generated by symmetry, providing a direct numerical test of the time-reversal relations discussed in the main text.

\begin{table}[h!]
\caption{Pump and probe parameters.}
\label{tab:pulses}
\begin{ruledtabular}
\begin{tabular}{lcc}
                    & pump                 & probe \\
\colrule
 $\hbar\omega$ (eV) & 1.69                 & 37.5 \\
 duration (fs)      & 7.2                  & 0.40 \\
 $I_0$ (W/cm$^2$)   & $10^{10}$, $10^{11}$ & $10^{8}$ \\
\end{tabular}
\end{ruledtabular}
\end{table}

The probe-induced current is obtained by subtracting the pump-only current, $\Delta\mathbf{j}=\mathbf{j}_{\mathrm{pump+probe}}-\mathbf{j}_{\mathrm{pump}}$. Since the simulations are finite in time, both $\Delta\mathbf{j}$ and the probe field are multiplied for $t>\tau$ by $e^{-\Gamma(t-\tau)}$, with $\Gamma^{-1}=\fs{20}$, prior to Fourier transformation. This windowing is distinct from the physical dephasing time $\Ttwo$. The helicity-resolved conductivity is $\sigma_{\pm}(\omega,\tau)=\Delta\tilde{j}_{\pm}/\Delta\tilde{E}_{\pm}$, where $j_{\pm}=(j_x\mp ij_y)/\sqrt{2}$ is the $\sigma^{\pm}$ helicity component (and $E_{\pm}$ likewise), the tilde denotes the windowed Fourier transform, and $\Delta\mathbf{E}=\mathbf{E}_{\mathrm{pump+probe}}-\mathbf{E}_{\mathrm{pump}}$ is the probe field. Following Ref.~\cite{wu2016theory}, replacing the atomic dipole by the electronic current,
\begin{equation}
  \varepsilon_{\pm}=1+\frac{4\pi i\,\alpha\,\sigma_{\pm}}{\omega},
  \quad
  \mu_{\pm}=\frac{2\omega}{c}\,\mathrm{Im}\sqrt{\varepsilon_{\pm}} .
  \label{eq:mu}
\end{equation}
The pump-induced differential absorption,
\[
\Delta\mu_{\pm}=\mu_{\pm}(\omega,\tau)-\mu_{\pm}^{\mathrm{probe\;only}}(\omega),
\]
is the quantity plotted throughout. Here $\alpha$ is an effective thickness that converts the two-dimensional sheet conductivity into a three-dimensional dielectric response. We fix $\alpha$ by requiring $\max|4\pi\alpha\sigma_{\pm}/\omega|=0.1$, which keeps the film optically thin, so that $\mu_{\pm}$ is linear in $\sigma_{\pm}$ and line shapes are undistorted. $\alpha$ therefore sets only the overall scale of $\Delta\mu_{\pm}$, while line shapes, helicity contrasts, and all normalized observables are independent of this convention. Separately, doubling the probe-field amplitude leaves the extracted $\sigma_{\pm}$ unchanged, so the probe operates in the linear-response regime.

\section{Appendix B: Selection rules at \texorpdfstring{$K$ and \Kp{}}{K and K'}}
\label{app:selection}

The asymmetries listed in Table~\ref{tab:rules} originate from the crystal symmetry at $K$, whose little group is $C_{3h}$. Because the vertical mirrors of the full point group exchange $K$ and \Kp{} rather than constraining either valley individually, the states are classified by their $C_3$ crystal angular momentum $m$ (mod~3) and horizontal-mirror parity $\sigma_h$, listed in Table~\ref{tab:irreps}. Electric-dipole transitions conserve both quantum numbers~\cite{geondzhian2022valley,cao2012valley,rana2022generation,rana2022probing},
\begin{equation}
m_f=m_i\pm1 \!\!\pmod 3,
\qquad
\sigma_h(f)\,\sigma_h(\varepsilon)\,\sigma_h(i)=+1 .
\label{eq:rules}
\end{equation}
The three Mo $4p$ core states ($p_\pm$ and $p_z$) span only \ev{0.16}, comparable to the dephasing-limited linewidth $2\hbar/\Ttwo=\ev{0.13}$, so they are unresolved and contribute as a single absorption edge.
\begin{table}[h!]
\caption{Symmetry labels at $K$. Rows 1--2 are the Mo $4p$ core states, rows 3--4 the band states, row 5 the dipole operator.}
\label{tab:irreps}
\begin{ruledtabular}
\begin{tabular}{lccc}
            & irrep & $m$ & $\sigma_h$ \\
\colrule
$p_{\pm}$  & $E'$  & $\pm1$ & $+$ \\
$p_z$      & $A''$ & $0$    & $-$ \\
\dpm{}     & $E'$  & $\mp1$ & $+$ \\
$d_{z^2}$  & $A'$  & $0$    & $+$ \\
$x\pm iy$  & $E'$  & $\pm1$ & $+$ \\
\end{tabular}
\end{ruledtabular}
\end{table}

{\it (i) The conduction minimum is a symmetry-protected null.} The $A'$ conduction state $d_{z^2}$ is reached from $p_{-}$ by \sigp{} and from $p_{+}$ by \sigm{}. Both routes involve the same reduced matrix element of the on-site Mo $4p\to4d$ transition, giving $|D_{+}|=|D_{-}|$ and therefore $\eta=0$ at both valleys~\cite{geondzhian2022valley}. In the crystal the $p_{\pm}$ states are not exactly equivalent. Their weak in-plane dispersion splits them by \ev{0.07} and the two dipole amplitudes differ by 4\%, giving a computed asymmetry $\eta=+0.02$ rather than an exact zero~\cite{SM}. The energy splitting also displaces the two absorption channels spectrally, so the $d_0$ line carries a very weak dispersive CD. Odd in energy, it largely cancels in the band-integrated dichroism.

{\it (ii) The \dpm{} channels are locked to one helicity.} At $K$, the valence-band maximum $d_{+2}$ is reached from $p_{+}$ by \sigp{}. The \sigm{} channel would instead require the $p_z$ core state, whose odd mirror parity forbids the transition, $(+)(+)(-)=-1$. Consequently, the core$\rightarrow$valence transition absorbs only \sigp{} at $K$ and only \sigm{} at \Kp{} by time reversal. Applying the same selection rules to the pump transition $d_{+2}\rightarrow d_{z^2}$ shows that the helicity that writes a valley is also the helicity that selectively probes its hole population. This is why the co-rotating probe in Fig.~\ref{fig:readout}(a) is the bright one.

At $K$, the \dpm{} transitions attain their symmetry-imposed limits, with $\eta=+1.00$ for the hole line (core$\to d_{+2}$) and $\eta=-1.00$ for the strongest electron line (core$\to d_{-2}$). This asymmetry specifies the helicity selectivity of a transition, but not its oscillator strength. Thus, although the transitions into CB$+1$ and CB$+3$ are also fully helicity selective, their dipole strengths are only about one third of that into CB$+2$. The stronger CB$+2$ transition reflects its mirror-even $d_{-2}$ character and the correspondingly large on-site Mo $4p\rightarrow4d$ dipole matrix element, and CB$+2$ therefore dominates the electron-channel response~\cite{liu2013three,geondzhian2022valley}. By contrast, the lowest $d_{z^2}$ conduction band is helicity insensitive. This null does not erase the valley information. Because optical excitation conserves the valley index, each photoexcited electron is correlated with a hole in the same valley, and a valley-selective measurement of the hole population therefore also identifies the valley population of the accompanying electrons.

\bibliography{references}

\end{document}